\documentclass[a4paper,fleqn,usenatbib]{mnras}

\usepackage[T1]{fontenc}
\usepackage{ae,aecompl}

\usepackage{graphicx}	% Including figure files
\usepackage{amsmath}	% Advanced maths commands
\usepackage{amssymb}	% Extra maths symbols
\usepackage{multirow}
\usepackage{epsfig}
\usepackage{subfigure}
\usepackage{newtxtext,newtxmath}

\title[X-ray polarization in MCG-05-23-16]{Polarization constraints on the X-ray corona in Seyfert Galaxies: MCG-05-23-16}

\author[{\it IXPE} Collaboration]{A. Marinucci,$^{1}\thanks{E-mail: andrea.marinucci@asi.it (AM)}$
F. Muleri,$^{2}$
M. Dovčiak,$^{3}$
S. Bianchi,$^{4}$
F. Marin,$^{5}$
G. Matt,$^{4}$
F. Ursini,$^{4}$
\newauthor
R. Middei,$^{6,7}$
H. L. Marshall,$^{8}$
L. Baldini,$^{9,10}$
T. Barnouin,$^{5}$
N. Cavero Rodriguez,$^{11}$
\newauthor
A. De Rosa,$^{2}$
L. Di Gesu,$^{1}$
D. Harper,$^{11}$
A. Ingram,$^{12}$
V. Karas,$^{3}$
H. Krawczynski,$^{11}$
\newauthor
G. Madejski,$^{13}$
C. Panagiotou,$^{8}$
P. O. Petrucci,$^{14}$
J. Podgorny,$^{5,3,15}$
S. Puccetti,$^{6}$
\newauthor
F. Tombesi,$^{16,17,18}$
A. Veledina,$^{19,20,21}$
W. Zhang,$^{22}$
I. Agudo,$^{23}$
L. A. Antonelli,$^{7,6}$
\newauthor
M. Bachetti,$^{24}$
W. H. Baumgartner,$^{25}$
R. Bellazzini,$^{9}$
S. D. Bongiorno,$^{25}$
R. Bonino,$^{26,27}$
\newauthor
A. Brez,$^{9}$
N. Bucciantini,$^{28,29,30}$
F. Capitanio,$^{2}$
S. Castellano,$^{9}$
E. Cavazzuti,$^{1}$
\newauthor
S. Ciprini,$^{17,6}$
E. Costa,$^{2}$
E. Del Monte,$^{2}$
N. Di Lalla,$^{13}$
A. Di Marco,$^{2}$
I. Donnarumma,$^{1}$
\newauthor
V. Doroshenko,$^{31,21}$
S. R. Ehlert,$^{25}$
T. Enoto,$^{32}$
Y. Evangelista,$^{2}$
S. Fabiani,$^{2}$
\newauthor
R. Ferrazzoli,$^{2}$
J. A. Garcia,$^{33}$
S. Gunji,$^{34}$
K. Hayashida,$^{35}$
J. Heyl,$^{36}$
W. Iwakiri,$^{37}$
\newauthor
S. G. Jorstad,$^{38,39}$
T. Kitaguchi,$^{32}$
J. J. Kolodziejczak,$^{25}$
F. La Monaca,$^{2}$
L. Latronico,$^{26}$
\newauthor
I. Liodakis,$^{40}$
S. Maldera,$^{26}$
A. Manfreda,$^{9}$
A. P. Marscher,$^{38}$
I. Mitsuishi,$^{41}$
T. Mizuno,$^{42}$
\newauthor
C.-Y. Ng,$^{43}$
S. L. O'Dell,$^{25}$
N. Omodei,$^{13}$
C. Oppedisano,$^{26}$
A. Papitto,$^{7}$
G. G. Pavlov,$^{44}$
\newauthor
A. L. Peirson,$^{13}$
M. Perri,$^{6,7}$
M. Pesce-Rollins,$^{9}$
M. Pilia,$^{24}$
A. Possenti,$^{24}$
\newauthor
J. Poutanen,$^{19,21}$
B. D. Ramsey,$^{25}$
J. Rankin,$^{2}$
A. Ratheesh,$^{2}$
R. W. Romani,$^{13}$
C. Sgrò,$^{9}$
\newauthor
P. Slane,$^{45}$
P. Soffitta,$^{2}$
G. Spandre,$^{9}$
T. Tamagawa,$^{32}$
F. Tavecchio,$^{46}$
R. Taverna,$^{47}$
\newauthor
Y. Tawara,$^{41}$
A. F. Tennant,$^{25}$
N. E. Thomas,$^{25}$
A. Trois,$^{24}$
S. S. Tsygankov,$^{19,21}$
\newauthor
R. Turolla,$^{47,48}$
J. Vink,$^{49}$
M. C. Weisskopf,$^{25}$
K. Wu,$^{48}$
F. Xie,$^{50, 2}$
S. Zane$^{48}$
}

\date{Accepted XXX. Received YYY; in original form ZZZ}

\pubyear{2022}

\begin{document}
\label{firstpage}
\pagerange{\pageref{firstpage}--\pageref{lastpage}}
\maketitle

\begin{abstract}
\indent We report on the first observation of a radio-quiet Active Galactic Nucleus (AGN) using polarized X-rays: the Seyfert 1.9 galaxy MCG-05-23-16. This source was pointed with the {\it Imaging X-ray Polarimetry Explorer} ({\it IXPE}) starting on May 14, 2022 for a net observing time of 486 ks, simultaneously with XMM-{\it Newton} (58 ks) and {\it NuSTAR} (83 ks). A polarization degree smaller than $\Pi<4.7\%$ (at the 99\% c.l.) is derived in the 2-8 keV energy range, where emission is dominated by the primary component ascribed to the hot corona. The broad-band spectrum, inferred from a simultaneous fit to the {\it IXPE}, {\it NuSTAR}, and XMM-{\it Newton} data, is well reproduced by a power law with photon index $\Gamma=1.85\pm0.01$ and a high-energy cutoff $E_{\rm C}=120\pm15$ keV. A comparison with Monte Carlo simulations shows that a lamp-post and a conical geometry of the corona are consistent with the observed upper limit, a slab geometry is allowed only if the inclination angle of the system is less than 50$^{\circ}$.
\end{abstract}

\begin{keywords}
Galaxies: active - Galaxies: Seyfert - Individual: MCG-05-23-16
\end{keywords}

%%%%%%%%%%%%%%%%%%%%%%%%%%%%%%%%%%%%%%%%%%%%%%%%%%

%%%%%%%%%%%%%%%%% BODY OF PAPER %%%%%%%%%%%%%%%%%%
\section{Introduction}
It is now widely accepted that the primary X-ray emission of Seyfert galaxies is produced by multiple up-scattering events of cool photons by hot electrons: the Comptonization process \citep{st80, zpj00}. However, the energy supply of this medium and the conditions leading to a formation of the hot plasma close to the black hole are debated. The physical picture of plasma fuelling through the gravitational energy transformation greatly depends on the geometry and size of the hot medium.  In one scenario, the energy dissipation (and electron heating) is distributed over a large volume, with characteristic sizes $\sim10-100\ R_{\rm g}$ (where $R_g=GM/c^2$ is the gravitational radius, $G$ is the gravitational constant, $M$ is black hole mass and $c$ is the speed of light). Early studies considered the so-called \textit{slab-corona} model, where the hot medium was assumed to be distributed above the cold accretion disc \citep{hm91, hm93}, and presumably energized by some disc instability, likely of magnetic origin \citep{merloni03}. On the other hand, in a \textit{lamp-post} geometry, the primary X-ray emission is assumed to be coming from a compact source ($\sim1-10\ R_{\rm g}$), located on the accretion disc axis \citep{flb17} and could be associated with an aborted jet \citep{ghm04}.

Spectroscopic analyses have in principle the capability to constrain the coronal geometry but, even the best available observations
provided by {\it NuSTAR}, while good enough to measure the physical coronal parameters like the optical depth and the temperature, are not able to distinguish statistically among different geometries \citep{tbm18, mbm19}.

A very promising and powerful tool to assess the coronal geometry is reverberation mapping of the corona-disc system \citep[][and references therein]{ucf14}.
In fact, the disc response to the corona illumination depends also on the geometry of the latter \citep{wcf16}. However, to fully exploit this technique, observations with the next generation X-ray observatories such as {\it eXTP} and {\it Athena} are required \citep{derosa19, dmb13}, even if very long XMM-$Newton$ observations can already deliver some results \citep{fac17}.

\indent X-ray polarization provides an independent tool to constrain the coronal geometry. Polarization, in fact, is extremely sensitive to the geometry of the emitting matter and of the photon field \citep{sk10,bkm17,tmb18,zdb19}. With the aim to constrain its coronal geometry, the {\it Imaging X-ray Polarimetry Explorer} \citep[{\it IXPE}:][]{w16} observed the bright Seyfert galaxy MCG-05-23-16. 

\indent MCG-05-23-16 is a nearby ($z$=0.0085 or 36 Mpc, \citealt{wbw03}) Seyfert 1.9 galaxy with broad emission lines in the near-infrared \citep{Goodrich1994}. It has been extensively observed in X-rays \citep{wym97,bbm04,brd07,rad07,per02, bcg08,mbm13}, showing a moderate cold absorption ($N_H \sim$10$^{22}$ cm$^{-2}$).  Recently,
{\it NuSTAR} observations were able to constrain the high energy cutoff ($E_{\rm C} \sim100-160$ keV, variable on a time scale of $\sim$100 ks) and therefore the coronal physical parameters kT$_e$ and $\tau$ \citep{bmh15, zmm17}. 
X-ray reverberation features have also been detected with XMM-{\it Newton} in this source \citep{zrc13,kaf16}.\\
\indent In the optical and near-infrared wavelengths (0.4$-$2.2$\mu$m) the source exhibits a low continuum linear polarization degree (1$-$2\%) and  a polarized flux density which increases with wavelength, a possible sign of Compton scattering or a different non-thermal component at work \citep{bhb90}.

With a 2$-$10 keV flux of (7$-$10)$\times10^{-11}$ erg cm$^{-2}$ s$^{-1}$ \citep{mw04}, MCG-05-23-16 is one of the brightest Seyfert galaxies, only moderately variable on both short and long time-scales, with a relatively simple spectrum (no significant absorption in the {\it IXPE} band) and well measured coronal parameters. It is therefore the ideal target to search for polarization signatures of the coronal geometry in radio-quiet AGN. The paper is organized as follows: in Sect. 2 we discuss the data reduction procedure, while in Sect. 3  we present the spectro-polarimetric analyses. Our results are then discussed and summarized in Sect. 4 and 5.

\section{Observations and data reduction}
\label{reduction}
{\it IXPE} \citep{ixpe} observed MCG-05-23-16 starting on May 14, 2022 with its three Detector Units (DU), for a net exposure time of 47 ks. The pointing of the source started again on May 21, for additional 439 ks. 
Cleaned level 2 event files were produced and calibrated using standard filtering criteria with the dedicated {\sc ftools} tasks and the latest calibration files available in the {\it IXPE} calibration database (CALDB 20211118). $I$, $Q$, $U$ Stokes background spectra were extracted from source-free circular regions with a radius of 100 arcsec. Extraction radii for the $I$ Stokes spectra of the source were computed via an iterative process which leads to the maximization of the Signal-to-Noise Ratio (SNR) in the 2-8 keV energy band, similar to the approach described in \citet{pico04}. We therefore adopted  circular regions centered on the source with radii of 62 arcsec, 57 arcsec and 62 arcsec for DU1, DU2 and DU3, respectively. The net exposure times are 485.7 ks and the same extraction radii were then applied to the $Q$ and $U$ Stokes spectra. We used a constant energy binning of 0.2 keV for $Q$, $U$ Stokes spectra and required a SNR higher than 5 in each spectral channel, in the intensity spectra. $I$, $Q$, $U$ Stokes spectra from the three DUs are always fitted independently in the following, but we will plot them together using the {\sc setp group} command in {\sc Xspec}, for the sake of visual clarity. Background represents the 2.0\%, 1.8\% and 2.1\% of the total DU1, DU2 and DU3 $I$ spectra, respectively.  The summed background subtracted light curves show an average count rate $C_{\rm 2-8\ keV}$=0.525$\pm$0.002 cts s$^{-1}$ with a level of variability of $\sim$20$-$30\%, in the range 0.33-0.79 cts s$^{-1}$.

XMM-{\it Newton} started its observation on May 21, 2022 for 83 ks of elapsed time with the EPIC CCD cameras: the pn \citep{struder01} and the two MOS \citep{turner01}, operated in small window and medium filter mode. Data from the MOS detectors are not included in our analysis due to pile-up. The data from the pn camera show no significant pile-up as indicated by the {\sc epatplot} output. The extraction radii and the optimal time cuts for flaring particle background were computed with SAS 20 \citep{gabr04} with the same SNR maximization procedure reported above. The resulting optimal extraction radii for the source and the background spectra are 40 and 50 arcsec, respectively. The net exposure time for the pn time-averaged spectrum is 58.1 ks.

{\it NuSTAR} \citep{nustar} observed MCG-05-23-16 simultaneously to XMM-{\it Newton}, with its two coaligned X-ray telescopes with corresponding Focal Plane Module A (FPMA) and B (FPMB). The total elapsed time is 171.4 ks. The Level 1 data products were processed with the {\it NuSTAR} Data Analysis Software (NuSTARDAS) package (v. 2.1.2). Cleaned event files (level 2 data products) were produced and calibrated using standard filtering criteria with the \textsc{nupipeline} task and the latest calibration files available in the {\it NuSTAR} calibration database (CALDB 20220510). Extraction radii for the source and background spectra were $40$ arcsec and 60 arcsec, FPMA spectra were binned in order not to over-sample the instrumental resolution more than a factor of 2.5 and to have a SNR  greater than 5 in each spectral channel, the same energy binning was then applied to the FPMB spectra. The net observing
times for the FPMA and the FPMB data sets are 83.4 ks and 83 ks, respectively.\\
We adopt the cosmological parameters $H_0=70$ km s$^{-1}$ Mpc$^{-1}$, $\Omega_\Lambda=0.73$ and $\Omega_{\rm m}=0.27$, i.e. the default ones in \textsc{Xspec 12.12.1} \citep{Xspec}. Errors correspond to the 90\% confidence level for one interesting parameter ($\Delta\chi^2=2.7$), if not stated otherwise. 

\begin{figure}
  \epsfig{file=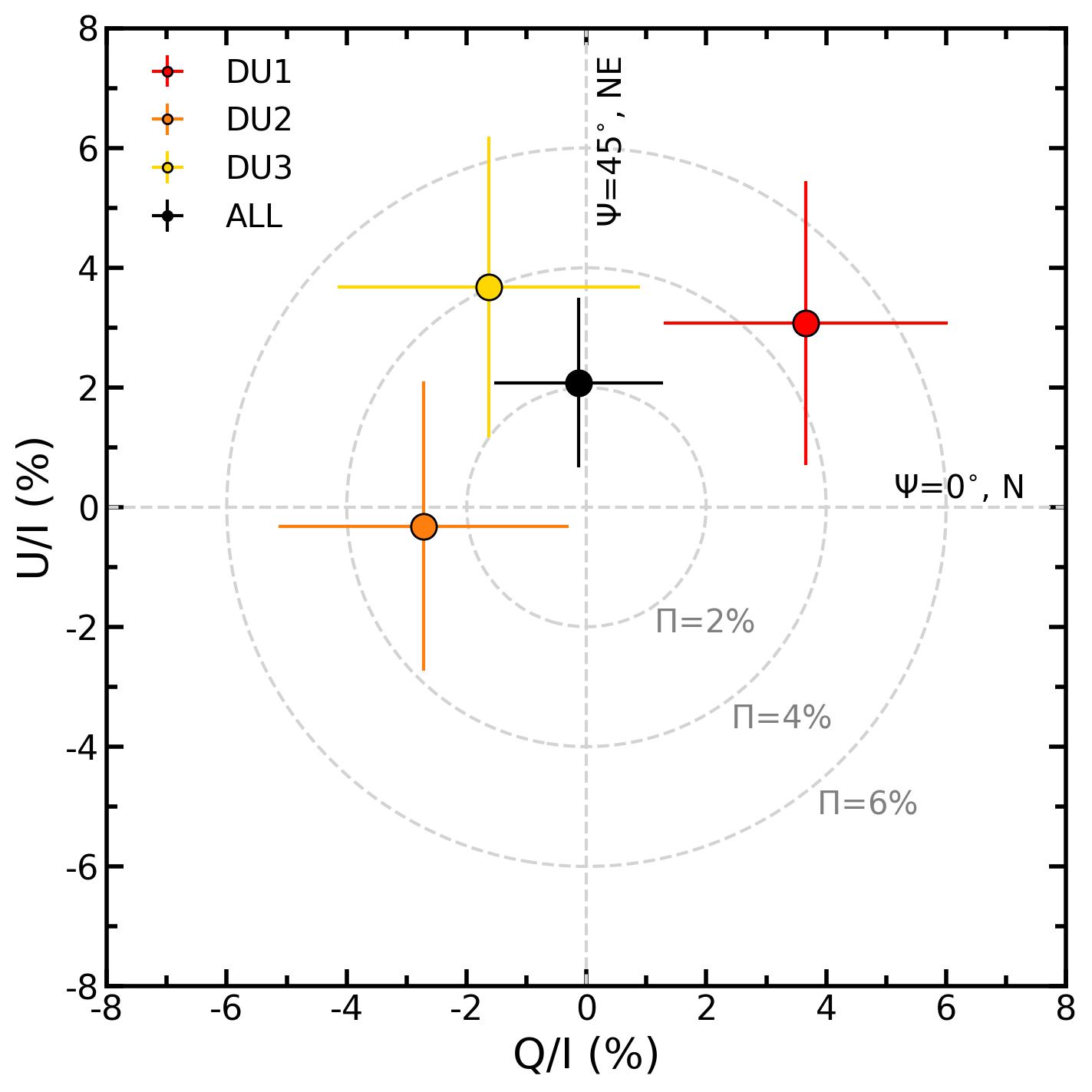, width=1.0\columnwidth}
    \vspace{-0.4cm}
  \caption{Normalized $U/I$ and $Q/I$ Stokes parameters are shown, calculated using the full 2-8 keV {\it IXPE} band. Uncertainties are reported at the 68\% c.l.}
  \label{stokes}
\end{figure}

\begin{figure*}
 \epsfig{file=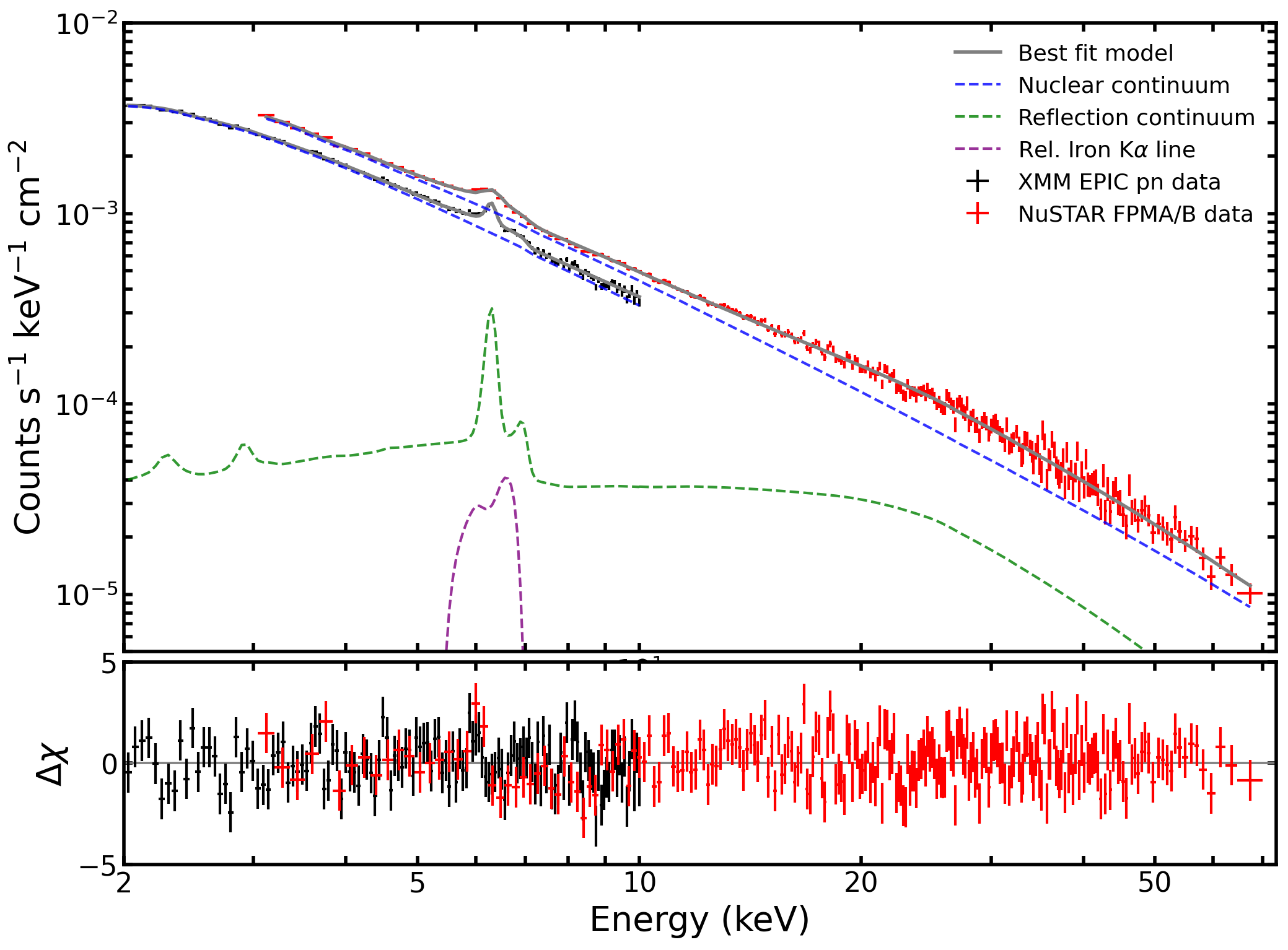, width=0.96\columnwidth}
  \epsfig{file=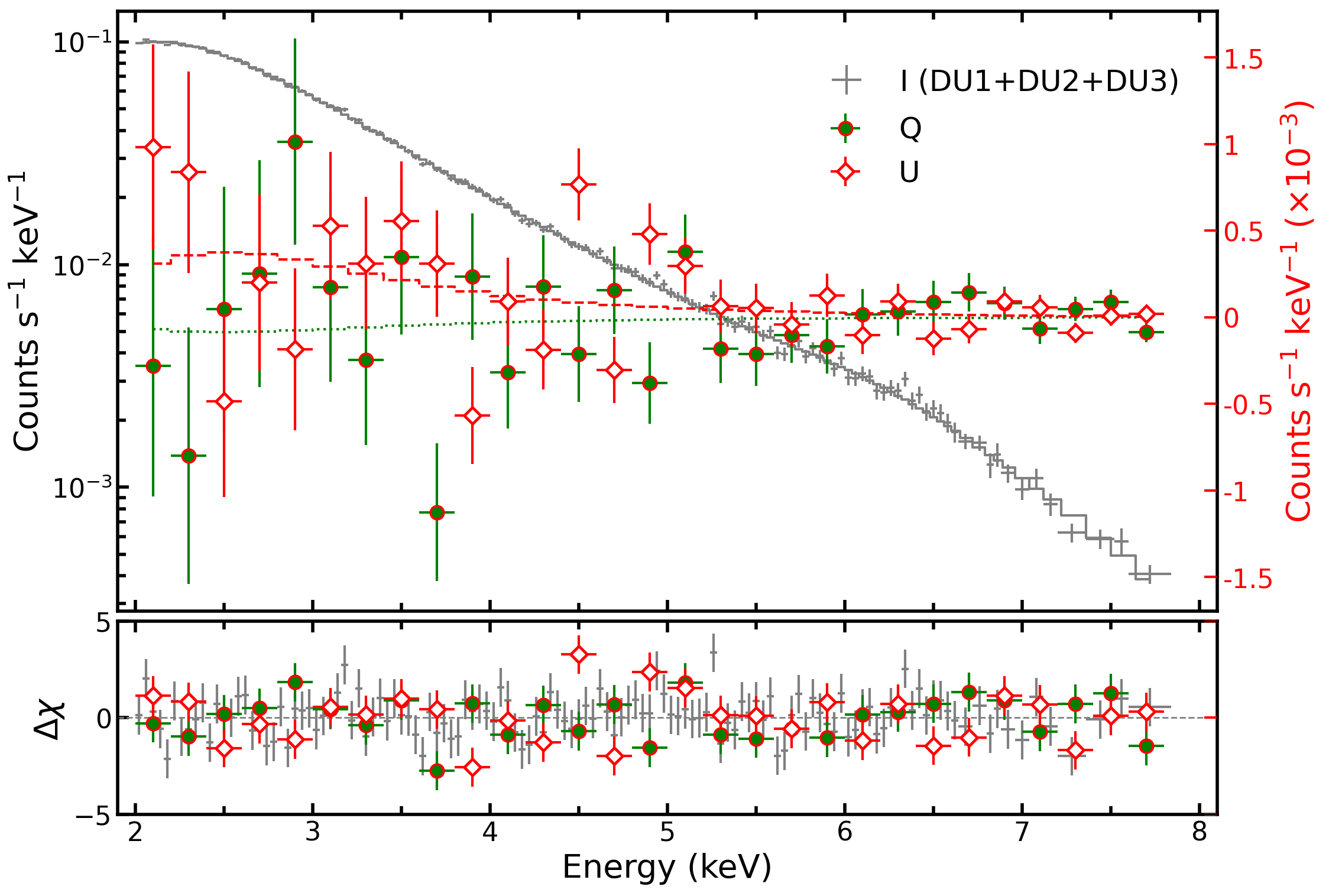, width=1.04\columnwidth}
  \caption{{\it Left panel:} The simultaneous  EPIC pn and the grouped {\it NuSTAR} FPMA and FPMB spectra of MCG-05-23-16 divided by the relative effective area are shown with residuals. The best fitting model is shown as a solid grey line and the different components as dashed lines. {\it Right panel:} {\it IXPE} $I$ (grey circles), $Q$ (green circles) and $U$ (red empty circles) grouped Stokes spectra are shown with residuals, along with the corresponding best fitting models. Note the different scales on the $
  y$-axes for $I$ and $Q$/$U$ data.}
  \label{mcg5_fit}
\end{figure*}
\vspace{-0.2cm}
\section{Data analysis}
\subsection{{\it IXPE} polarimetric analysis}
We start investigating the polarized signal from MCG-05-23-16 by analyzing its polarization cubes, which are the simplest data structures holding polarization information. They can be created using {\sc ixpeobssim} \citep[version 26.3.3:][]{baldini22}. This applies the \citet{kcb15} formalism to a user-defined set of events to compute the Stokes parameters, the Minimum Detectable Polarization \citep[MDP:][]{eow12}, the polarization degree, the polarization angle, and the associated uncertainties. In our case, we created one polarization cube for each DU and then one combining the three, using the whole 2$-$8 keV band.
Fig. \ref{stokes} shows the normalized $U/I$ and $Q/I$ Stokes parameters, data counts from background regions have been subtracted. Using the polarization cube from the three combined DUs, we retrieve a MDP=4.3\%, a polarization degree $\Pi=(2.1\pm1.4)\%$ and angle $\Psi=(47\pm19)^{\circ}$ (using 68\% c.l. on one single parameter). We do not constrain any energy dependence of the polarization properties.

\begin{table}
\begin{center}
\begin{tabular}{llll}
\multicolumn{2}{c}{\bf Parameter} & \multicolumn{2}{c}{\bf Best fitting values}  \\
\hline
 \multicolumn{4}{c}{\sc cutoff power law}  \\
\multicolumn{2}{l}{ N$_{\rm H}$ (cm$^{-2}$)}&   \multicolumn{2}{l}{$(1.35\pm0.05)\times10^{22}$}  \\
\multicolumn{2}{l}{$\Gamma$}&  \multicolumn{2}{l}{$1.85\pm0.01$}    \\
\multicolumn{2}{l}{$E_{\rm C}$ (keV)} & \multicolumn{2}{l}{$120\pm15$} \\ 
\multicolumn{2}{l}{$N$} & \multicolumn{2}{l}{$(2.60\pm0.05)\times10^{-2}$}   \\
\multicolumn{2}{l}{$\Pi$ (\%)}& \multicolumn{2}{l}{$2.2\pm1.7$}\\
\multicolumn{2}{l}{$\Psi$ ($^{\circ}$)} & \multicolumn{2}{l}{$50\pm25$} \\ 
    \multicolumn{4}{c}{\sc xillver} \\
\multicolumn{2}{l}{$R$} & \multicolumn{2}{l}{$0.30\pm0.05$}    \\
\multicolumn{2}{l}{$N$} &   \multicolumn{2}{l}{$2.14_{-0.1}^{+0.4}\times10^{-2}$} \\
\multicolumn{2}{l}{$\Pi$ (\%)}& \multicolumn{2}{l}{-}\\
\multicolumn{2}{l}{$\Psi$ ($^{\circ}$)} & \multicolumn{2}{l}{-} \\ 
    \multicolumn{4}{c}{\sc kerrdisk} \\
\multicolumn{2}{l}{$\theta$ ($^{\circ}$)} &   \multicolumn{2}{l}{$48_{-8}^{+12}$} \\
\multicolumn{2}{l}{$N$} & \multicolumn{2}{l}{$(3.9\pm_{-0.5}^{+0.8})\times10^{-5}$}    \\
\multicolumn{2}{l}{$\Pi$ (\%)}& \multicolumn{2}{l}{-}\\
\multicolumn{2}{l}{$\Psi$ ($^{\circ}$)} & \multicolumn{2}{l}{-} \\ 
 & & & \\
\multicolumn{2}{l}{$\chi^2$/dof} & \multicolumn{2}{l}{1250/1169} \\
 \multicolumn{2}{l}{F$_{2-10}$ (erg cm$^{-2}$ s$^{-1}$)} & \multicolumn{2}{l}{$(7.45\pm0.05)\times10^{-11}$} \\
\multicolumn{2}{l}{L$_{2-10}$ (erg s$^{-1}$)} & \multicolumn{2}{l}{$(1.20\pm0.02)\times10^{43}$}\\
\multicolumn{4}{c}{\sc cross-calibrations} \\
\multicolumn{2}{l}{\sc constants} & \multicolumn{2}{l}{\sc gain} \\ 
$C_{\rm pn-DU1}$ & $1.09_{-0.01}^{+0.02}$ & $\alpha_{\rm DU1}$& $0.953\pm0.009$ \\
$C_{\rm pn-DU2}$ & $1.06\pm0.02$ & $\beta_{\rm DU1}$& $0.07\pm0.03$ \\
$C_{\rm pn-DU3}$ & $0.97\pm0.02$ & $\alpha_{\rm DU2}$& $0.963_{-0.009}^{+0.007}$\\
$C_{\rm pn-FPMA}$ & $1.39\pm0.01$& $\beta_{\rm DU2}$& $0.04\pm0.03$ \\
$C_{\rm pn-FPMB}$ & $1.43\pm0.01$ & $\alpha_{\rm DU3}$& $0.951_{-0.007}^{+0.009}$\\
 &  & $\beta_{\rm DU3}$& $0.07_{-0.04}^{+0.03}$\\
\hline
\end{tabular}
\end{center}
\caption{\label{best_par_mcg} Best fit parameters from the joint fit. Normalization units are in photons keV$^{-1}$ cm$^{-2}$ s$^{-1}$. $R$ is the reflection fraction measured as the ratio between the Compton reflection and the primary component fluxes between 20 and 40 keV. The 2$-$10 keV flux is retrieved from the EPIC-pn data.} 
\end{table}

\subsection{XMM-{\it Newton}, {\it NuSTAR} and {\it IXPE} spectro-polarimetric analysis}
We started modelling the simultaneous 2$-$10 keV XMM-{\it Newton} and 3$-$79 keV {\it NuSTAR} spectra of MCG-05-23-16 with a model composed of an absorbed cutoff power law ({\sc zTBabs $\times$ cutoffpl} in {\sc Xspec}) and a Compton reflection component \citep[{\sc xillver}:][]{garcia13}. The former reproduces the primary continuum of the source while the latter takes into account reflection off neutral, distant material. Galactic absorption is modeled with {\sc TBabs}, using a column density N$_{\rm H}=7.8\times10^{20}$ cm$^{-2}$ \citep{hi4pi} and multiplicative constants take into account cross-calibration uncertainties between the FPMA, the FPMB and EPIC pn. The 
photon index and cutoff energy of the reflection continuum is linked to the one of the primary continuum, iron abundance is fixed to the solar one and the inclination angle to $\theta=30^{\circ}$. The resulting $\chi^2$/dof is good (785/628) but some residuals appear at $\sim$6 keV\footnote{The inferred energy of the Fe K$\alpha$ line in the pn spectrum is not consistent with 6.4 keV and we therefore added a {\sc vashift} component in the model. We retrieve $v=2230_{-550}^{+510}$ km s$^{-1}$. Since this effect is not found in the MOS spectra we conclude that is likely due to calibration issues in the pn.}. This could be indicative of a second iron K$\alpha$ component, smeared by relativistic effects in the inner regions of the accretion disc. Indeed, when compared with old XMM data, residuals are perfectly consistent with the ones presented in \citet{brd07}. A further spectral component is therefore included: \citep[{\sc kerrdisk:}][]{br06}. The black hole spin is fixed to $a=0.998$, the emissivity to $\epsilon(r)=r^{-3}$, the rest-frame energy of the emission line at 6.4 keV and the inner radius of the disc to R$_{\rm in}=37$ R$_{\rm g}$ \citep[as reported in the simultaneous XMM+Suzaku analysis:][]{rad07}. We obtain a best fit $\chi^2$/dof=683/625  (Fig. \ref{mcg5_fit}) and an inclination angle $\theta=48_{-8}^{+12}\degr$. A more detailed analysis of the complete data set will be presented in a forthcoming paper.

We then included {\it IXPE} $I$ Stokes spectra to the XMM and {\it NuSTAR} fit. We followed the formalism discussed in \citet{stro17} and used the weighted analysis method presented in \citet{dimarco22} (parameter {\sc stokes=Neff} in {\sc xselect}). We obtain a $\chi^2$/dof=1378/1000 due to presence of large residuals at the low and high energies in the {\it IXPE} $I$ spectra. This has already been observed in other bright sources and can be likely explained in terms of calibration issues \citep{taverna22, kmd22}. We therefore modified the response files gains in the $I$ spectra (using {\sc gain fit} command) and obtained a $\chi^2$/dof=1055/994. We then included the $Q$ and $U$ Stokes spectra and linked their gain parameters to the ones of the $I$ spectra. Cross-calibration constants are included and the three spectral components of the model are convolved with the polarization model {\sc polconst}. Two parameters can be then be inferred for each spectral component: the polarization degree $\Pi$ and angle $\Psi$, both constant functions of the energy. The polarization degree and angle associated to the nuclear continuum are left free to vary while the ones associated to the other spectral components are fixed to $\Pi=0\%$ and $\Psi=0^{\circ}$. We will verify a posteriori that the fit is insensitive to these values. We retrieve the photon index $\Gamma=1.85\pm0.01$ and the cutoff energy $E_{\rm C}=120\pm15$ keV, in agreement with previously reported values \citep[][]{bmh15, zmm17}. The best spectro-polarimetric joint fit provides the polarization degree $\Pi$=$(2.2\pm1.7)$\% and $\Psi$=$(50\pm25)^{\circ}$ for the nuclear power law. Therefore, at the 99\% c.l. (for one single parameter of interest, $\Delta \chi^2=6.63$) we only get an upper limit $\Pi=4.7\%$. The contour plots between these two parameters are shown in Fig. \ref{mcg5_cplot}. If the cutoff power law is substituted by the Comptonization model {\sc compps} \citep{ps96} an electron temperature kT$_e$=25$\pm2$ keV and optical depth $\tau=1.27\pm0.08$ are retrieved, assuming a slab geometry ($\chi^2$/dof=1248/1169).  \\
\indent On theoretical grounds, the Compton reflection continuum is expected to be moderately polarized but the iron K$\alpha$ fluorescence emission line arising from the same scattering material is not \citep[as shown in][]{gm11, marin18}. We also note that scattering from distant AGN components do not significantly impact the measured polarization due to their low contribution in this energy band \citep{Marin2018b}. However, we also tried to leave the polarization degree of the {\sc xillver} and {\sc kerrdisk} components free to vary in the fit. No statistically significant improvement is found ($\chi^2$/dof=1250/1169) and the fit is insensitive to these two parameters. 
\vspace{-0.2 cm}
\section{Discussion}
\begin{figure}
\vspace{-0.8cm}
\centering
\epsfig{file=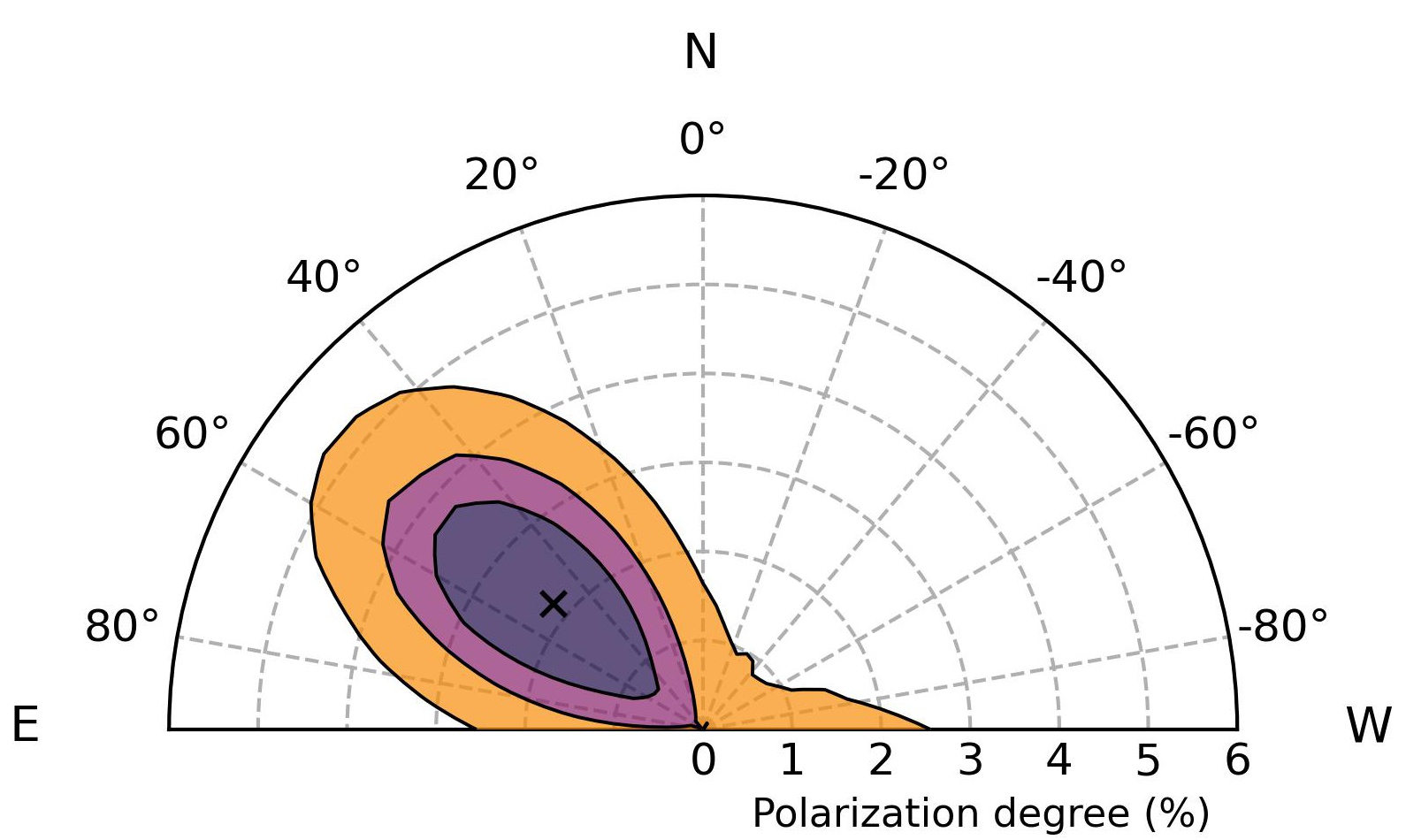, width=1.0\columnwidth}
\epsfig{file=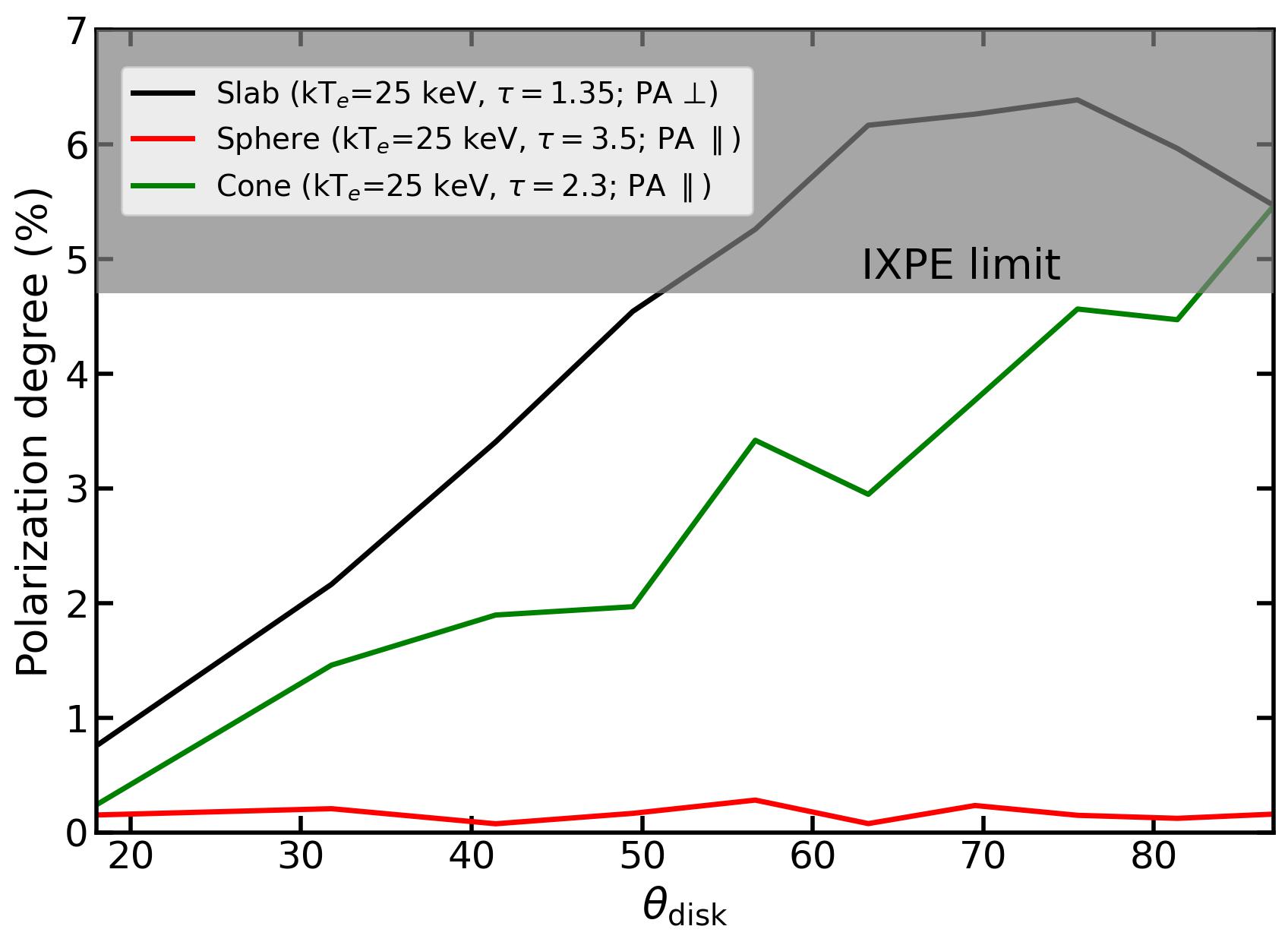, width=0.9\columnwidth}
  \caption{{\it Top-panel:} contour plot between the polarization degree $\Pi$ and angle $\Psi$ associated to the primary power law component. Purple, pink and orange shaded regions indicate 68\%, 90\% and 99\% confidence levels for two parameters of interest, respectively. {\it Bottom-panel:} Monte Carlo simulations performed with the Comptonization code {\sc monk}. The input coronal parameters kT$_e$, $\tau$ correspond to the spectral shape reported in Tab. \ref{best_par_mcg} and the polarization angle is reported with respect to the accretion disc plane. }
  \label{mcg5_cplot}
\end{figure}

\indent Several radio-loud AGN have been observed in the first {\it IXPE} months of operations and a highly significant polarized signal has been measured in two blazars so far, due to synchrotron emission in the jet. On the other hand, polarized X-rays produced in radio quiet sources are thought to be produced by the inverse Compton mechanism, occurring within tens of gravitational radii from the central black hole. Different geometries of the scattering medium,  depending on the degree of asymmetry, will result in a different degree of polarization. \\
\indent MCG-05-23-16 is the first radio quiet AGN observed by {\it IXPE} and the nuclear power law component, ascribed to the hot corona, contributes to the 94$\pm$3\%  of the 2-8 keV flux.  We found that the best spectro-polarimetric fit of the simultaneous {\it IXPE}-{\it NuSTAR}-XMM data provides only, at the 99\% c.l., an upper limit $\Pi=4.7\%$.
Three different geometries for the hot corona in this AGN have been recently explored in \citet{umb22}, with the Comptonization code {\sc monk} \citep{zdb19}. Fig. \ref{mcg5_cplot} shows the results with input coronal temperatures kT$_e$=25 keV and different Thomson optical depths $\tau$, which correspond to the spectral shape of the continuum reported in Tab. \ref{best_par_mcg}. 
The three coronal geometries are: a slab sandwitching the accretion disc, a spherical lamp-post on the symmetry axis, and a truncated cone in outflow to describe the base of a failed jet. 
%\citet{umb22} found that a slab corona yields a polarization degree $\Pi$=0-6.5\%, a spherical corona yields $\Pi<0.3\%$ and a conical corona yields $\Pi$=0.3-5.5\%, as function of the inclination angle of the disc.
The geometrical parameters for the simulations shown in Fig. \ref{mcg5_cplot} are as follows. The slab corona fully covers a truncated disc with an inner radius of 30 $R_g$. The spherical lamp-post has a radius of 10 $R_g$ and a height above the disc of 30 $R_g$. As for the truncated cone, the distance between the lower base and the disc and the vertical thickness are both of 20 $R_g$, while the semi-aperture is 30$^{\circ}$, and the outflow velocity is $0.3c$. Among these geometries, the slab corona is the only one that gives a polarization vector perpendicular to the disc plane, the other two predicting it to be parallel to the accretion disc. We note that the polarization angle for the three geometries does not significantly vary for different input kT$_e$ and $\tau$ values, while the absolute value of the polarization degree can change. However, this does not impact much on the relative differences between the three geometries.

VLA observations at 8.4 GHz of the source showed a possible elongation in PA$\simeq169^{\circ}$ east of north \citep{mfn09} while Hubble Space Telescope WFPC2 images revealed {\sc [O iii]} emission in PA$\simeq40^{\circ}$ \citep{fpa00}. 
The 90\% c.l. contour plot shows that the polarization angle is roughly aligned with the {\sc [O iii]} emitting region, which likely traces the Narrow Line Region (NLR) in this object. Assuming that the accretion disc is perpendicular to the NLR, a slab corona geometry  would produce such a polarization angle, similar to what observed in the X-ray binary Cyg X-1 \citep{kmd22}. Of course, given the low significance of this result, it must be taken as no more than suggestive, needing more data for a confirmation. Highly inclined slab geometries (with $\theta\gtrsim50^{\circ}$) can be ruled out and our estimate of  $\theta=48_{-8}^{+12}\degr$, from the broad iron K$\alpha$ profile, does not provide further constraints on the hot coronal geometry. 
\vspace{-0.2cm}
\section{Conclusions}
The launch of {\it IXPE}, on December 9, 2021, opened a new observing window on the study of supermassive black holes and MCG-05-23-16 is the perfect candidate to investigate the hot corona close to the supermassive black hole.  Simultaneously observed with {\it IXPE}, XMM-{\it Newton} and {\it NuSTAR} in May 2022, the source showed a low level of neutral absorption along the line of sight (N$_{\rm H}=1.35\pm0.05\times10^{22}$ cm$^{-2}$), a modest level of Compton reflection ($R=0.30\pm0.05$) and broad relativistic iron K$\alpha$ emission line. This translates into a power-law continuum which largely dominates the total flux in the 2$-$8 keV {\it IXPE} band ($F_{\rm pow}/F_{\rm tot}=94\pm3\%$) and when it is convolved with a constant polarization model we obtain a 99\% c.l. upper limit to the polarization fraction of 4.7\%. This result is consistent with a spherical ("lamp-post") and a conical geometry of the corona, while for a slab corona (with kT$_e$=25 keV and $\tau=1.35$) it implies an inclination angle less than 50$^{\circ}$.
\vspace{-0.2cm}
\section*{Acknowledgements}
The {\it Imaging X ray Polarimetry Explorer} ({\it IXPE}) is a joint US and Italian mission.  The US contribution is supported by the National Aeronautics and Space Administration (NASA) and led and managed by its Marshall Space Flight Center (MSFC), with industry partner Ball Aerospace (contract NNM15AA18C).  The Italian contribution is supported by the Italian Space Agency (ASI) through contract ASI-OHBI-2017-12-I.0, agreements ASI-INAF-2017-12-H0 and ASI-INFN-2017.13-H0, and its Space Science Data Center (SSDC), and by the Istituto Nazionale di Astrofisica (INAF) and the Istituto Nazionale di Fisica Nucleare (INFN) in Italy.  This research used data products provided by the {\it IXPE} Team (MSFC, SSDC, INAF, and INFN) and distributed with additional software tools by the High-Energy Astrophysics Science Archive Research Center (HEASARC), at NASA Goddard Space Flight Center (GSFC). Part of the French contribution is supported by the Scientific Research National Center (CNRS) and the French spatial agency (CNES).

%%%%%%%%%%%%%%%%%%%%%%%%%%%%%%%%%%%%%%%%%%%%%%%%%%



%\bibliographystyle{mnras}

\bibliographystyle{mnras}
\bibliography{sbs} 

\begin{thebibliography}{}
\makeatletter
\relax
\def\mn@urlcharsother{\let\do\@makeother \do\$\do\&\do\#\do\^\do\_\do\%\do\~}
\def\mn@doi{\begingroup\mn@urlcharsother \@ifnextchar [ {\mn@doi@}
  {\mn@doi@[]}}
\def\mn@doi@[#1]#2{\def\@tempa{#1}\ifx\@tempa\@empty \href
  {http://dx.doi.org/#2} {doi:#2}\else \href {http://dx.doi.org/#2} {#1}\fi
  \endgroup}
\def\mn@eprint#1#2{\mn@eprint@#1:#2::\@nil}
\def\mn@eprint@arXiv#1{\href {http://arxiv.org/abs/#1} {{\tt arXiv:#1}}}
\def\mn@eprint@dblp#1{\href {http://dblp.uni-trier.de/rec/bibtex/#1.xml}
  {dblp:#1}}
\def\mn@eprint@#1:#2:#3:#4\@nil{\def\@tempa {#1}\def\@tempb {#2}\def\@tempc
  {#3}\ifx \@tempc \@empty \let \@tempc \@tempb \let \@tempb \@tempa \fi \ifx
  \@tempb \@empty \def\@tempb {arXiv}\fi \@ifundefined
  {mn@eprint@\@tempb}{\@tempb:\@tempc}{\expandafter \expandafter \csname
  mn@eprint@\@tempb\endcsname \expandafter{\@tempc}}}

\bibitem[\protect\citeauthoryear{{Arnaud}}{{Arnaud}}{1996}]{Xspec}
{Arnaud} K.~A.,  1996, in ASP Conf. Ser. 101: Astronomical Data Analysis
  Software and Systems V. p.~17

\bibitem[\protect\citeauthoryear{{Baldini}, {Bucciantini}, {Di Lalla},
  {Ehlert}, {Manfreda}, {Omodei}, {Pesce-Rollins}  \& {Sgr{\`o}}}{{Baldini}
  et~al.}{2022}]{baldini22}
{Baldini} L.,  {Bucciantini} N.,  {Di Lalla} N.,  {Ehlert} S.~R.,  {Manfreda}
  A.,  {Omodei} N.,  {Pesce-Rollins} M.,   {Sgr{\`o}} C.,  2022, arXiv
  e-prints, \href {https://ui.adsabs.harvard.edu/abs/2022arXiv220306384B} {p.
  arXiv:2203.06384}

\bibitem[\protect\citeauthoryear{{Balestra}, {Bianchi}  \& {Matt}}{{Balestra}
  et~al.}{2004}]{bbm04}
{Balestra} I.,  {Bianchi} S.,   {Matt} G.,  2004, \mn@doi [\aap]
  {10.1051/0004-6361:20030211}, \href
  {https://ui.adsabs.harvard.edu/abs/2004A&A...415..437B} {415, 437}

\bibitem[\protect\citeauthoryear{{Balokovi{\'c}} et~al.,}{{Balokovi{\'c}}
  et~al.}{2015}]{bmh15}
{Balokovi{\'c}} M.,  et~al., 2015, \mn@doi [\apj] {10.1088/0004-637X/800/1/62},
  \href {https://ui.adsabs.harvard.edu/abs/2015ApJ...800...62B} {800, 62}

\bibitem[\protect\citeauthoryear{{Beckmann}, {Courvoisier}, {Gehrels},
  {Lubi{\'n}ski}, {Malzac}, {Petrucci}, {Shrader}  \& {Soldi}}{{Beckmann}
  et~al.}{2008}]{bcg08}
{Beckmann} V.,  {Courvoisier} T.~J.~L.,  {Gehrels} N.,  {Lubi{\'n}ski} P.,
  {Malzac} J.,  {Petrucci} P.~O.,  {Shrader} C.~R.,   {Soldi} S.,  2008,
  \mn@doi [\aap] {10.1051/0004-6361:200810674}, \href
  {https://ui.adsabs.harvard.edu/abs/2008A&A...492...93B} {492, 93}

\bibitem[\protect\citeauthoryear{{Beheshtipour}, {Krawczynski}  \&
  {Malzac}}{{Beheshtipour} et~al.}{2017}]{bkm17}
{Beheshtipour} B.,  {Krawczynski} H.,   {Malzac} J.,  2017, \mn@doi [\apj]
  {10.3847/1538-4357/aa906a}, \href
  {http://adsabs.harvard.edu/abs/2017ApJ...850...14B} {850, 14}

\bibitem[\protect\citeauthoryear{{Braito} et~al.,}{{Braito}
  et~al.}{2007}]{brd07}
{Braito} V.,  et~al., 2007, \mn@doi [\apj] {10.1086/521916}, \href
  {https://ui.adsabs.harvard.edu/abs/2007ApJ...670..978B} {670, 978}

\bibitem[\protect\citeauthoryear{{Brenneman} \& {Reynolds}}{{Brenneman} \&
  {Reynolds}}{2006}]{br06}
{Brenneman} L.~W.,  {Reynolds} C.~S.,  2006, \mn@doi [\apj] {10.1086/508146},
  \href {https://ui.adsabs.harvard.edu/abs/2006ApJ...652.1028B} {652, 1028}

\bibitem[\protect\citeauthoryear{{Brindle}, {Hough}, {Bailey}, {Axon}, {Ward},
  {Sparks}  \& {McLean}}{{Brindle} et~al.}{1990}]{bhb90}
{Brindle} C.,  {Hough} J.~H.,  {Bailey} J.~A.,  {Axon} D.~J.,  {Ward} M.~J.,
  {Sparks} W.~B.,   {McLean} I.~S.,  1990, \mnras, \href
  {https://ui.adsabs.harvard.edu/abs/1990MNRAS.244..604B} {244, 604}

\bibitem[\protect\citeauthoryear{{De Rosa} et~al.,}{{De Rosa}
  et~al.}{2019}]{derosa19}
{De Rosa} A.,  et~al., 2019, \mn@doi [Science China Physics, Mechanics, and
  Astronomy] {10.1007/s11433-018-9297-0}, \href
  {https://ui.adsabs.harvard.edu/abs/2019SCPMA..6229504D} {62, 29504}

\bibitem[\protect\citeauthoryear{{Di Marco} et~al.,}{{Di Marco}
  et~al.}{2022}]{dimarco22}
{Di Marco} A.,  et~al., 2022, \mn@doi [\aj] {10.3847/1538-3881/ac51c9}, \href
  {https://ui.adsabs.harvard.edu/abs/2022AJ....163..170D} {163, 170}

\bibitem[\protect\citeauthoryear{{Dovciak} et~al.,}{{Dovciak}
  et~al.}{2013}]{dmb13}
{Dovciak} M.,  et~al., 2013, arXiv e-prints, \href
  {https://ui.adsabs.harvard.edu/abs/2013arXiv1306.2331D} {p. arXiv:1306.2331}

\bibitem[\protect\citeauthoryear{{Elsner}, {O'Dell}  \& {Weisskopf}}{{Elsner}
  et~al.}{2012}]{eow12}
{Elsner} R.~F.,  {O'Dell} S.~L.,   {Weisskopf} M.~C.,  2012, in Space
  Telescopes and Instrumentation 2012: Ultraviolet to Gamma Ray. p. 84434N
  (\mn@eprint {arXiv} {1208.0610}), \mn@doi{10.1117/12.924889}

\bibitem[\protect\citeauthoryear{{Fabian}, {Alston}, {Cackett}, {Kara},
  {Uttley}  \& {Wilkins}}{{Fabian} et~al.}{2017a}]{fac17}
{Fabian} A.~C.,  {Alston} W.~N.,  {Cackett} E.~M.,  {Kara} E.,  {Uttley} P.,
  {Wilkins} D.~R.,  2017a, \mn@doi [Astronomische Nachrichten]
  {10.1002/asna.201713341}, \href
  {https://ui.adsabs.harvard.edu/abs/2017AN....338..269F} {338, 269}

\bibitem[\protect\citeauthoryear{{Fabian}, {Lohfink}, {Belmont}, {Malzac}  \&
  {Coppi}}{{Fabian} et~al.}{2017b}]{flb17}
{Fabian} A.~C.,  {Lohfink} A.,  {Belmont} R.,  {Malzac} J.,   {Coppi} P.,
  2017b, \mn@doi [\mnras] {10.1093/mnras/stx221}, \href
  {http://adsabs.harvard.edu/abs/2017MNRAS.467.2566F} {467, 2566}

\bibitem[\protect\citeauthoryear{{Ferruit}, {Wilson}  \& {Mulchaey}}{{Ferruit}
  et~al.}{2000}]{fpa00}
{Ferruit} P.,  {Wilson} A.~S.,   {Mulchaey} J.,  2000, \mn@doi [\apjs]
  {10.1086/313379}, \href
  {https://ui.adsabs.harvard.edu/abs/2000ApJS..128..139F} {128, 139}

\bibitem[\protect\citeauthoryear{{Gabriel} et~al.,}{{Gabriel}
  et~al.}{2004}]{gabr04}
{Gabriel} C.,  et~al., 2004, in {F.~Ochsenbein, M.~G.~Allen, \& D.~Egret} ed.,
  Astronomical Society of the Pacific Conference Series Vol. 314, Astronomical
  Data Analysis Software and Systems (ADASS) XIII. pp 759--+

\bibitem[\protect\citeauthoryear{{Garc{\'{\i}}a}, {Dauser}, {Reynolds},
  {Kallman}, {McClintock}, {Wilms}  \& {Eikmann}}{{Garc{\'{\i}}a}
  et~al.}{2013}]{garcia13}
{Garc{\'{\i}}a} J.,  {Dauser} T.,  {Reynolds} C.~S.,  {Kallman} T.~R.,
  {McClintock} J.~E.,  {Wilms} J.,   {Eikmann} W.,  2013, \mn@doi [\apj]
  {10.1088/0004-637X/768/2/146}, \href
  {http://adsabs.harvard.edu/abs/2013ApJ...768..146G} {768, 146}

\bibitem[\protect\citeauthoryear{{Ghisellini}, {Haardt}  \&
  {Matt}}{{Ghisellini} et~al.}{2004}]{ghm04}
{Ghisellini} G.,  {Haardt} F.,   {Matt} G.,  2004, \mn@doi [\aap]
  {10.1051/0004-6361:20031562}, \href
  {https://ui.adsabs.harvard.edu/abs/2004A&A...413..535G} {413, 535}

\bibitem[\protect\citeauthoryear{{Goodrich}, {Veilleux}  \& {Hill}}{{Goodrich}
  et~al.}{1994}]{Goodrich1994}
{Goodrich} R.~W.,  {Veilleux} S.,   {Hill} G.~J.,  1994, \mn@doi [\apj]
  {10.1086/173746}, \href
  {https://ui.adsabs.harvard.edu/abs/1994ApJ...422..521G} {422, 521}

\bibitem[\protect\citeauthoryear{{Goosmann} \& {Matt}}{{Goosmann} \&
  {Matt}}{2011}]{gm11}
{Goosmann} R.~W.,  {Matt} G.,  2011, \mn@doi [\mnras]
  {10.1111/j.1365-2966.2011.18923.x}, \href
  {https://ui.adsabs.harvard.edu/abs/2011MNRAS.415.3119G} {415, 3119}

\bibitem[\protect\citeauthoryear{{HI4PI Collaboration} et~al.,}{{HI4PI
  Collaboration} et~al.}{2016}]{hi4pi}
{HI4PI Collaboration} et~al., 2016, \mn@doi [\aap]
  {10.1051/0004-6361/201629178}, \href
  {https://ui.adsabs.harvard.edu/abs/2016A&A...594A.116H} {594, A116}

\bibitem[\protect\citeauthoryear{{Haardt} \& {Maraschi}}{{Haardt} \&
  {Maraschi}}{1991}]{hm91}
{Haardt} F.,  {Maraschi} L.,  1991, \apjl, \href
  {http://cdsads.u-strasbg.fr/cgi-bin/nph-bib_query?bibcode=1991ApJ...380L..51H&amp;db_key=AST}
  {380, L51}

\bibitem[\protect\citeauthoryear{{Haardt} \& {Maraschi}}{{Haardt} \&
  {Maraschi}}{1993}]{hm93}
{Haardt} F.,  {Maraschi} L.,  1993, \mn@doi [\apj] {10.1086/173020}, \href
  {https://ui.adsabs.harvard.edu/abs/1993ApJ...413..507H} {413, 507}

\bibitem[\protect\citeauthoryear{{Harrison} et~al.,}{{Harrison}
  et~al.}{2013}]{nustar}
{Harrison} F.~A.,  et~al., 2013, \mn@doi [\apj] {10.1088/0004-637X/770/2/103},
  \href {http://adsabs.harvard.edu/abs/2013ApJ...770..103H} {770, 103}

\bibitem[\protect\citeauthoryear{{Kara}, {Alston}, {Fabian}, {Cackett},
  {Uttley}, {Reynolds}  \& {Zoghbi}}{{Kara} et~al.}{2016}]{kaf16}
{Kara} E.,  {Alston} W.~N.,  {Fabian} A.~C.,  {Cackett} E.~M.,  {Uttley} P.,
  {Reynolds} C.~S.,   {Zoghbi} A.,  2016, \mn@doi [\mnras]
  {10.1093/mnras/stw1695}, \href
  {https://ui.adsabs.harvard.edu/abs/2016MNRAS.462..511K} {462, 511}

\bibitem[\protect\citeauthoryear{{Kislat}, {Clark}, {Beilicke}  \&
  {Krawczynski}}{{Kislat} et~al.}{2015}]{kcb15}
{Kislat} F.,  {Clark} B.,  {Beilicke} M.,   {Krawczynski} H.,  2015, \mn@doi
  [Astroparticle Physics] {10.1016/j.astropartphys.2015.02.007}, \href
  {https://ui.adsabs.harvard.edu/abs/2015APh....68...45K} {68, 45}

\bibitem[\protect\citeauthoryear{{Krawczynski} et~al.,}{{Krawczynski}
  et~al.}{2022}]{kmd22}
{Krawczynski} H.,  et~al., 2022, arXiv e-prints, \href
  {https://ui.adsabs.harvard.edu/abs/2022arXiv220609972K} {p. arXiv:2206.09972}

\bibitem[\protect\citeauthoryear{{Marin}}{{Marin}}{2018}]{marin18}
{Marin} F.,  2018, \mn@doi [\aap] {10.1051/0004-6361/201833225}, \href
  {https://ui.adsabs.harvard.edu/abs/2018A&A...615A.171M} {615, A171}

\bibitem[\protect\citeauthoryear{{Marin}, {Dov{\v{c}}iak}  \&
  {Kammoun}}{{Marin} et~al.}{2018}]{Marin2018b}
{Marin} F.,  {Dov{\v{c}}iak} M.,   {Kammoun} E.~S.,  2018, \mn@doi [\mnras]
  {10.1093/mnras/sty1062}, \href
  {https://ui.adsabs.harvard.edu/abs/2018MNRAS.478..950M} {478, 950}

\bibitem[\protect\citeauthoryear{{Mattson} \& {Weaver}}{{Mattson} \&
  {Weaver}}{2004}]{mw04}
{Mattson} B.~J.,  {Weaver} K.~A.,  2004, \mn@doi [\apj] {10.1086/380502}, \href
  {https://ui.adsabs.harvard.edu/abs/2004ApJ...601..771M} {601, 771}

\bibitem[\protect\citeauthoryear{{Merloni}}{{Merloni}}{2003}]{merloni03}
{Merloni} A.,  2003, \mn@doi [\mnras] {10.1046/j.1365-8711.2003.06496.x}, \href
  {https://ui.adsabs.harvard.edu/abs/2003MNRAS.341.1051M} {341, 1051}

\bibitem[\protect\citeauthoryear{{Middei}, {Bianchi}, {Marinucci}, {Matt},
  {Petrucci}, {Tamborra}  \& {Tortosa}}{{Middei} et~al.}{2019}]{mbm19}
{Middei} R.,  {Bianchi} S.,  {Marinucci} A.,  {Matt} G.,  {Petrucci} P.~O.,
  {Tamborra} F.,   {Tortosa} A.,  2019, \mn@doi [\aap]
  {10.1051/0004-6361/201935881}, \href
  {https://ui.adsabs.harvard.edu/abs/2019A&A...630A.131M} {630, A131}

\bibitem[\protect\citeauthoryear{{Molina}, {Bassani}, {Malizia}, {Stephen},
  {Bird}, {Bazzano}  \& {Ubertini}}{{Molina} et~al.}{2013}]{mbm13}
{Molina} M.,  {Bassani} L.,  {Malizia} A.,  {Stephen} J.~B.,  {Bird} A.~J.,
  {Bazzano} A.,   {Ubertini} P.,  2013, \mn@doi [\mnras]
  {10.1093/mnras/stt844}, \href
  {http://adsabs.harvard.edu/abs/2013MNRAS.433.1687M} {433, 1687}

\bibitem[\protect\citeauthoryear{{Mundell}, {Ferruit}, {Nagar}  \&
  {Wilson}}{{Mundell} et~al.}{2009}]{mfn09}
{Mundell} C.~G.,  {Ferruit} P.,  {Nagar} N.,   {Wilson} A.~S.,  2009, \mn@doi
  [\apj] {10.1088/0004-637X/703/1/802}, \href
  {https://ui.adsabs.harvard.edu/abs/2009ApJ...703..802M} {703, 802}

\bibitem[\protect\citeauthoryear{{Perola}, {Matt}, {Cappi}, {Fiore},
  {Guainazzi}, {Maraschi}, {Petrucci}  \& {Piro}}{{Perola}
  et~al.}{2002}]{per02}
{Perola} G.~C.,  {Matt} G.,  {Cappi} M.,  {Fiore} F.,  {Guainazzi} M.,
  {Maraschi} L.,  {Petrucci} P.~O.,   {Piro} L.,  2002, \aap, 389, 802

\bibitem[\protect\citeauthoryear{{Piconcelli}, {Jimenez-Bail{\' o}n},
  {Guainazzi}, {Schartel}, {Rodr{\'{\i}}guez-Pascual}  \& {Santos-Lle{\'
  o}}}{{Piconcelli} et~al.}{2004}]{pico04}
{Piconcelli} E.,  {Jimenez-Bail{\' o}n} E.,  {Guainazzi} M.,  {Schartel} N.,
  {Rodr{\'{\i}}guez-Pascual} P.~M.,   {Santos-Lle{\' o}} M.,  2004, \mnras,
  \href
  {http://esoads.eso.org/cgi-bin/nph-bib_query?bibcode=2004MNRAS.351..161P&amp;db_key=AST}
  {351, 161}

\bibitem[\protect\citeauthoryear{{Poutanen} \& {Svensson}}{{Poutanen} \&
  {Svensson}}{1996}]{ps96}
{Poutanen} J.,  {Svensson} R.,  1996, \mn@doi [\apj] {10.1086/177865}, \href
  {http://adsabs.harvard.edu/abs/1996ApJ...470..249P} {470, 249}

\bibitem[\protect\citeauthoryear{{Reeves} et~al.,}{{Reeves}
  et~al.}{2007}]{rad07}
{Reeves} J.~N.,  et~al., 2007, \mn@doi [\pasj] {10.1093/pasj/59.sp1.S301},
  \href {https://ui.adsabs.harvard.edu/abs/2007PASJ...59S.301R} {59, 301}

\bibitem[\protect\citeauthoryear{{Schnittman} \& {Krolik}}{{Schnittman} \&
  {Krolik}}{2010}]{sk10}
{Schnittman} J.~D.,  {Krolik} J.~H.,  2010, \mn@doi [\apj]
  {10.1088/0004-637X/712/2/908}, \href
  {http://adsabs.harvard.edu/abs/2010ApJ...712..908S} {712, 908}

\bibitem[\protect\citeauthoryear{{Strohmayer}}{{Strohmayer}}{2017}]{stro17}
{Strohmayer} T.~E.,  2017, \mn@doi [\apj] {10.3847/1538-4357/aa643d}, \href
  {https://ui.adsabs.harvard.edu/abs/2017ApJ...838...72S} {838, 72}

\bibitem[\protect\citeauthoryear{{Str{\"u}der} et~al.,}{{Str{\"u}der}
  et~al.}{2001}]{struder01}
{Str{\"u}der} L.,  et~al., 2001, \aap, \href
  {http://esoads.eso.org/cgi-bin/nph-bib_query?bibcode=2001A%26A...365L..18S&amp;db_key=AST}
  {365, L18}

\bibitem[\protect\citeauthoryear{{Sunyaev} \& {Titarchuk}}{{Sunyaev} \&
  {Titarchuk}}{1980}]{st80}
{Sunyaev} R.~A.,  {Titarchuk} L.~G.,  1980, \aap, \href
  {http://adsabs.harvard.edu/abs/1980A%26A....86..121S} {86, 121}

\bibitem[\protect\citeauthoryear{{Tamborra}, {Matt}, {Bianchi}  \&
  {Dov{\v{c}}iak}}{{Tamborra} et~al.}{2018}]{tmb18}
{Tamborra} F.,  {Matt} G.,  {Bianchi} S.,   {Dov{\v{c}}iak} M.,  2018, \mn@doi
  [\aap] {10.1051/0004-6361/201732023}, \href
  {https://ui.adsabs.harvard.edu/abs/2018A&A...619A.105T} {619, A105}

\bibitem[\protect\citeauthoryear{{Taverna} et~al.,}{{Taverna}
  et~al.}{2022}]{taverna22}
{Taverna} R.,  et~al., 2022, arXiv e-prints, \href
  {https://ui.adsabs.harvard.edu/abs/2022arXiv220508898T} {p. arXiv:2205.08898}

\bibitem[\protect\citeauthoryear{{Tortosa}, {Bianchi}, {Marinucci}, {Matt}  \&
  {Petrucci}}{{Tortosa} et~al.}{2018}]{tbm18}
{Tortosa} A.,  {Bianchi} S.,  {Marinucci} A.,  {Matt} G.,   {Petrucci} P.~O.,
  2018, \mn@doi [\aap] {10.1051/0004-6361/201732382}, \href
  {http://adsabs.harvard.edu/abs/2018A%26A...614A..37T} {614, A37}

\bibitem[\protect\citeauthoryear{{Turner} et~al.,}{{Turner}
  et~al.}{2001}]{turner01}
{Turner} M.~J.~L.,  et~al., 2001, \aap, \href
  {http://esoads.eso.org/cgi-bin/nph-bib_query?bibcode=2001A%26A...365L..27T&amp;db_key=AST}
  {365, L27}

\bibitem[\protect\citeauthoryear{{Ursini}, {Matt}, {Bianchi}, {Marinucci},
  {Dov{\v{c}}iak}  \& {Zhang}}{{Ursini} et~al.}{2022}]{umb22}
{Ursini} F.,  {Matt} G.,  {Bianchi} S.,  {Marinucci} A.,  {Dov{\v{c}}iak} M.,
  {Zhang} W.,  2022, \mn@doi [\mnras] {10.1093/mnras/stab3745}, \href
  {https://ui.adsabs.harvard.edu/abs/2022MNRAS.510.3674U} {510, 3674}

\bibitem[\protect\citeauthoryear{{Uttley}, {Cackett}, {Fabian}, {Kara}  \&
  {Wilkins}}{{Uttley} et~al.}{2014}]{ucf14}
{Uttley} P.,  {Cackett} E.~M.,  {Fabian} A.~C.,  {Kara} E.,   {Wilkins} D.~R.,
  2014, \mn@doi [\aapr] {10.1007/s00159-014-0072-0}, \href
  {https://ui.adsabs.harvard.edu/abs/2014A&ARv..22...72U} {22, 72}

\bibitem[\protect\citeauthoryear{{Weaver}, {Yaqoob}, {Mushotzky}, {Nousek},
  {Hayashi}  \& {Koyama}}{{Weaver} et~al.}{1997}]{wym97}
{Weaver} K.~A.,  {Yaqoob} T.,  {Mushotzky} R.~F.,  {Nousek} J.,  {Hayashi} I.,
   {Koyama} K.,  1997, \mn@doi [\apj] {10.1086/303488}, \href
  {https://ui.adsabs.harvard.edu/abs/1997ApJ...474..675W} {474, 675}

\bibitem[\protect\citeauthoryear{{Wegner} et~al.,}{{Wegner}
  et~al.}{2003}]{wbw03}
{Wegner} G.,  et~al., 2003, \mn@doi [\aj] {10.1086/378959}, \href
  {https://ui.adsabs.harvard.edu/abs/2003AJ....126.2268W} {126, 2268}

\bibitem[\protect\citeauthoryear{{Weisskopf} et~al.,}{{Weisskopf}
  et~al.}{2016}]{w16}
{Weisskopf} M.~C.,  et~al., 2016, \mn@doi [Results in Physics]
  {10.1016/j.rinp.2016.10.021}, \href
  {https://ui.adsabs.harvard.edu/abs/2016ResPh...6.1179W} {6, 1179}

\bibitem[\protect\citeauthoryear{{Weisskopf} et~al.,}{{Weisskopf}
  et~al.}{2022}]{ixpe}
{Weisskopf} M.~C.,  et~al., 2022, \mn@doi [Journal of Astronomical Telescopes,
  Instruments, and Systems] {10.1117/1.JATIS.8.2.026002}, \href
  {https://ui.adsabs.harvard.edu/abs/2022JATIS...8b6002W} {8, 026002}

\bibitem[\protect\citeauthoryear{{Wilkins}, {Cackett}, {Fabian}  \&
  {Reynolds}}{{Wilkins} et~al.}{2016}]{wcf16}
{Wilkins} D.~R.,  {Cackett} E.~M.,  {Fabian} A.~C.,   {Reynolds} C.~S.,  2016,
  \mn@doi [\mnras] {10.1093/mnras/stw276}, \href
  {https://ui.adsabs.harvard.edu/abs/2016MNRAS.458..200W} {458, 200}

\bibitem[\protect\citeauthoryear{{Zdziarski}, {Poutanen}  \&
  {Johnson}}{{Zdziarski} et~al.}{2000}]{zpj00}
{Zdziarski} A.~A.,  {Poutanen} J.,   {Johnson} W.~N.,  2000, \mn@doi [\apj]
  {10.1086/317046}, \href {http://adsabs.harvard.edu/abs/2000ApJ...542..703Z}
  {542, 703}

\bibitem[\protect\citeauthoryear{{Zhang}, {Dov{\v{c}}iak}  \& {Bursa}}{{Zhang}
  et~al.}{2019}]{zdb19}
{Zhang} W.,  {Dov{\v{c}}iak} M.,   {Bursa} M.,  2019, \mn@doi [\apj]
  {10.3847/1538-4357/ab1261}, \href
  {https://ui.adsabs.harvard.edu/abs/2019ApJ...875..148Z} {875, 148}

\bibitem[\protect\citeauthoryear{{Zoghbi}, {Reynolds}, {Cackett}, {Miniutti},
  {Kara}  \& {Fabian}}{{Zoghbi} et~al.}{2013}]{zrc13}
{Zoghbi} A.,  {Reynolds} C.,  {Cackett} E.~M.,  {Miniutti} G.,  {Kara} E.,
  {Fabian} A.~C.,  2013, \mn@doi [\apj] {10.1088/0004-637X/767/2/121}, \href
  {https://ui.adsabs.harvard.edu/abs/2013ApJ...767..121Z} {767, 121}

\bibitem[\protect\citeauthoryear{{Zoghbi} et~al.,}{{Zoghbi}
  et~al.}{2017}]{zmm17}
{Zoghbi} A.,  et~al., 2017, \mn@doi [\apj] {10.3847/1538-4357/aa582c}, \href
  {https://ui.adsabs.harvard.edu/abs/2017ApJ...836....2Z} {836, 2}

\makeatother
\end{thebibliography}

%%%%%%%%%%%%%%%%%%%%%%%%%%%%%%%%%%%%%%%%%%%%%%%%%%

%%%%%%%%%%%%%%%%% APPENDICES %%%%%%%%%%%%%%%%%%%%%

%\section{Best fit results with {\sc KYNrline}}


%%%%%%%%%%%%%%%%%%%%%%%%%%%%%%%%%%%%%%%%%%%%%%%%%%

% Don't change these lines
%\bsp	% typesetting comment
\label{lastpage}
\onecolumn
\noindent Affiliations: \\
\noindent
\normalsize{$^{1}$ASI - Agenzia Spaziale Italiana, Via del Politecnico snc, 00133 Roma, Italy}\\ 
\normalsize{$^{2}$INAF Istituto di Astrofisica e Planetologia Spaziali, Via del Fosso del Cavaliere 100, 00133 Roma, Italy}\\ 
\normalsize{$^{3}$Astronomical Institute of the Czech Academy of Sciences, Boční II 1401/1, 14100 Praha 4, Czech Republic}\\ 
\normalsize{$^{4}$Dipartimento di Matematica e Fisica, Universit\`a degli Studi Roma Tre, Via della Vasca Navale 84, 00146 Roma, Italy}\\ 
\normalsize{$^{5}$Université de Strasbourg, CNRS, Observatoire Astronomique de Strasbourg, UMR 7550, 67000 Strasbourg, France}\\ 
\normalsize{$^{6}$Space Science Data Center, Agenzia Spaziale Italiana, Via del Politecnico snc, 00133 Roma, Italy}\\ 
\normalsize{$^{7}$INAF Osservatorio Astronomico di Roma, Via Frascati 33, 00078 Monte Porzio Catone (RM), Italy}\\ 
\normalsize{$^{8}$MIT Kavli Institute for Astrophysics and Space Research, Massachusetts Institute of Technology, 77 Massachusetts Avenue, Cambridge, MA 02139, USA}\\ 
\normalsize{$^{9}$Istituto Nazionale di Fisica Nucleare, Sezione di Pisa, Largo B. Pontecorvo 3, 56127 Pisa, Italy}\\ 
\normalsize{$^{10}$Dipartimento di Fisica, Università di Pisa, Largo B. Pontecorvo 3, 56127 Pisa, Italy}\\ 
\normalsize{$^{11}$Physics Department and McDonnell Center for the Space Sciences, Washington University in St. Louis, St. Louis, MO 63130, USA}\\ 
\normalsize{$^{12}$School of Mathematics, Statistics, and Physics, Newcastle University, Newcastle upon Tyne NE1 7RU, UK}\\ 
\normalsize{$^{13}$Department of Physics and Kavli Institute for Particle Astrophysics and Cosmology, Stanford University, Stanford, California 94305, USA}\\ 
\normalsize{$^{14}$Université Grenoble Alpes, CNRS, IPAG, 38000 Grenoble, France}\\ 
\normalsize{$^{15}$Astronomical Institute, Charles University, V Holešovičkách 2, CZ-18000 Prague, Czech Republic}\\ 
\normalsize{$^{16}$Dipartimento di Fisica, Universit\`a degli Studi di Roma "Tor Vergata", Via della Ricerca Scientifica 1, 00133 Roma, Italy}\\ 
\normalsize{$^{17}$Istituto Nazionale di Fisica Nucleare, Sezione di Roma "Tor Vergata", Via della Ricerca Scientifica 1, 00133 Roma, Italy}\\ 
\normalsize{$^{18}$Department of Astronomy, University of Maryland, College Park, Maryland 20742, USA}\\ 
\normalsize{$^{19}$Department of Physics and Astronomy, 20014 University of Turku, Finland}\\ 
\normalsize{$^{20}$Nordita, KTH Royal Institute of Technology and Stockholm University, Hannes Alfv\'{e}ns v\"{a}g 12, SE-10691 Stockholm, Sweden}\\ 
\normalsize{$^{21}$Space Research Institute of the Russian Academy of Sciences, Profsoyuznaya Str. 84/32, Moscow 117997, Russia}\\ 
\normalsize{$^{22}$National Astronomical Observatories, Chinese Academy of Sciences, 20A Datun Road, Beijing 100101, China}\\ 
\normalsize{$^{23}$Instituto de Astrofísica de Andalucía—CSIC, Glorieta de la Astronomía s/n, 18008 Granada, Spain}\\ 
\normalsize{$^{24}$INAF Osservatorio Astronomico di Cagliari, Via della Scienza 5, 09047 Selargius (CA), Italy}\\ 
\normalsize{$^{25}$NASA Marshall Space Flight Center, Huntsville, AL 35812, USA}\\ 
\normalsize{$^{26}$Istituto Nazionale di Fisica Nucleare, Sezione di Torino, Via Pietro Giuria 1, 10125 Torino, Italy}\\ 
\normalsize{$^{27}$Dipartimento di Fisica, Università degli Studi di Torino, Via Pietro Giuria 1, 10125 Torino, Italy}\\ 
\normalsize{$^{28}$INAF Osservatorio Astrofisico di Arcetri, Largo Enrico Fermi 5, 50125 Firenze, Italy}\\ 
\normalsize{$^{29}$Dipartimento di Fisica e Astronomia, Università degli Studi di Firenze, Via Sansone 1, 50019 Sesto Fiorentino (FI), Italy}\\ 
\normalsize{$^{30}$Istituto Nazionale di Fisica Nucleare, Sezione di Firenze, Via Sansone 1, 50019 Sesto Fiorentino (FI), Italy}\\ 
\normalsize{$^{31}$Institut f\"ur Astronomie und Astrophysik, Universität Tübingen, Sand 1, 72076 T\"ubingen, Germany}\\ 
\normalsize{$^{32}$RIKEN Cluster for Pioneering Research, 2-1 Hirosawa, Wako, Saitama 351-0198, Japan}\\ 
\normalsize{$^{33}$California Institute of Technology, Pasadena, CA 91125, USA}\\ 
\normalsize{$^{34}$Yamagata University,1-4-12 Kojirakawa-machi, Yamagata-shi 990-8560, Japan}\\ 
\normalsize{$^{35}$Osaka University, 1-1 Yamadaoka, Suita, Osaka 565-0871, Japan}\\ 
\normalsize{$^{36}$University of British Columbia, Vancouver, BC V6T 1Z4, Canada}\\ 
\normalsize{$^{37}$Department of Physics, Faculty of Science and Engineering, Chuo University, 1-13-27 Kasuga, Bunkyo-ku, Tokyo 112-8551, Japan}\\ 
\normalsize{$^{38}$Institute for Astrophysical Research, Boston University, 725 Commonwealth Avenue, Boston, MA 02215, USA}\\ 
\normalsize{$^{39}$Department of Astrophysics, St. Petersburg State University, Universitetsky pr. 28, Petrodvoretz, 198504 St. Petersburg, Russia}\\ 
\normalsize{$^{40}$Finnish Centre for Astronomy with ESO,  20014 University of Turku, Finland}\\ 
\normalsize{$^{41}$Graduate School of Science, Division of Particle and Astrophysical Science, Nagoya University, Furo-cho, Chikusa-ku, Nagoya, Aichi 464-8602, Japan}\\ 
\normalsize{$^{42}$Hiroshima Astrophysical Science Center, Hiroshima University, 1-3-1 Kagamiyama, Higashi-Hiroshima, Hiroshima 739-8526, Japan}\\ 
\normalsize{$^{43}$Department of Physics, The University of Hong Kong, Pokfulam, Hong Kong}\\ 
\normalsize{$^{44}$Department of Astronomy and Astrophysics, Pennsylvania State University, University Park, PA 16802, USA}\\ 
\normalsize{$^{45}$Center for Astrophysics | Harvard \& Smithsonian, 60 Garden St, Cambridge, MA 02138, USA}\\ 
\normalsize{$^{46}$INAF Osservatorio Astronomico di Brera, Via E. Bianchi 46, 23807 Merate (LC), Italy}\\ 
\normalsize{$^{47}$Dipartimento di Fisica e Astronomia, Università degli Studi di Padova, Via Marzolo 8, 35131 Padova, Italy}\\ 
\normalsize{$^{48}$Mullard Space Science Laboratory, University College London, Holmbury St Mary, Dorking, Surrey RH5 6NT, UK}\\ 
\normalsize{$^{49}$Anton Pannekoek Institute for Astronomy \& GRAPPA, University of Amsterdam, Science Park 904, 1098 XH Amsterdam, The Netherlands}\\ 
\normalsize{$^{50}$Guangxi Key Laboratory for Relativistic Astrophysics, School of Physical Science and Technology, Guangxi University, Nanning 530004, China}\\ 
\\ 
\end{document}